\documentclass[conference]{IEEEtran}
\IEEEoverridecommandlockouts
\usepackage{cite}
\usepackage{amsmath,amssymb,amsfonts}

\usepackage{listings}
\usepackage{color}

\usepackage[english]{babel} 

\usepackage{algorithmic}
\usepackage{graphicx}
\usepackage{textcomp}
\def\BibTeX{{\rm B\kern-.05em{\sc i\kern-.025em b}\kern-.08em
    T\kern-.1667em\lower.7ex\hbox{E}\kern-.125emX}}

\definecolor{codegray}{rgb}{0.5,0.5,0.5}
\definecolor{backcolour}{rgb}{0.95,0.95,0.92}

\lstdefinestyle{mystyle}{
  backgroundcolor=\color{backcolour}, commentstyle=\color{codegreen},
  keywordstyle=\color{magenta},
  numberstyle=\tiny\color{codegray},
  stringstyle=\color{codepurple},
  basicstyle=\ttfamily\footnotesize,
  breakatwhitespace=false,         
  breaklines=true,                 
  captionpos=b,                    
  keepspaces=true,                 
  numbers=left,                    
  numbersep=5pt,                  
  showspaces=false,                
  showstringspaces=false,
  showtabs=false,                  
  tabsize=2
}

\begin{document}

\title{Variability-Aware Machine Learning Model Selection: Feature Modeling, Instantiation, and Experimental Case Study\\
}
\author{\IEEEauthorblockN{Cristina Tavares, Nathalia Nascimento, Paulo Alencar, Donald Cowan} 
\IEEEauthorblockA{\textit{David R. Cheriton School of Computer Science} \\
\textit{University of Waterloo}\\
Waterloo, Canada \\
\{cristina.tavares, nmoraesdonascimento, palencar, dcowan\}@uwaterloo.ca}}

\maketitle

\begin{abstract}
The emergence of machine learning (ML) has led to a transformative shift in software techniques and guidelines for building software applications that support data analysis process activities such as data ingestion, modeling, and deployment. 
Specifically, this shift is impacting ML model selection, which is one of the key phases in this process. Model selection is the process of selecting a model or a set of models for the analysis. There have been several advances in model selection from the standpoint of core ML methods, including basic probability measures and resampling methods. However, from a software engineering perspective, this selection is still an ad hoc and informal process, is not supported by a design approach and representation formalism that captures the selection process and can not support the specification of existing model selection procedures (e.g., heuristics). This selection also does not take into account in a transparent way and adapts to the variety of contextual factors that affect the model selection, such as data characteristics, number of features, prediction type, and their intricate dependencies. Futher, it is not interpretable in the sense of explaining why a model has been selected and does not take into account the contextual factors and their interdependencies in the experimental evaluation that leads to a specific technique selection.
In general, although the current literature provides a wide variety of ML techniques and algorithms, there is a lack of design approaches to support algorithm selection. 
In this paper, we present a variability-aware ML algorithm selection approach that takes into account the commonalities and variations in the model selection process. 
The applicability of the approach is illustrated by an experimental case study based on the Scikit-Learn heuristics, in which existing model selections presented in the literature are compared with selections suggested by the approach. 
The proposed approach can be seen as a step towards the provision of a more explicit, adaptive, transparent, interpretable, and automated basis for model selection. 
\end{abstract}


The emergence of machine learning (ML) has led to a transformative shift in software techniques and guidelines for building software applications that support data analysis process activities such as data ingestion, modeling, and deployment. Software projects based on ML have been increasingly developed in a wide range of application areas, including business, health, and commerce \cite{saltz2022current}. These projects typically need to support machine learning development processes that involve data, code, models, and numerous back-and-forth team interactions. These elements, as well as these complex feedback loops and interdependencies, make building software engineering solutions for ML applications more challenging than for traditional software applications. 

Model selection is the process of selecting an appropriate ML model or a set of models. The model selection process can be applied, for example, across different types of models in categories such as classification, regression, clustering, and dimensionality reduction. The selection of a classification model could involve models such as logistic regression (LR), support vector classifier (SVC), random forest (RF), and stochastic gradient descent (SGD), among many others.
This selection is often one of the most challenging tasks in machine learning application development. Not only does this selection require expert knowledge about the algorithms themselves, but also about numerous tacit factors that affect the selection. From an ML theoretical perspective, there have been several advances in model selection from the standpoint of core ML methods, including basic probabilistic measures and resampling methods. These models often limit their focus on in-sample and out-of-sample errors, respectively.

However, from a software engineering perspective, model selection is still an ad hoc and informal process, is not supported by a design approach and representation formalisms that capture the selection process, and can not support the specification of existing model selection procedures (e.g., heuristics). This selection also does not often take into account in a transparent way and adapts to the variety of contextual factors that affect the model selection, such as data characteristics, number of features, prediction type, and their intricate dependencies. It is not interpretable in the sense of explaining why a model has been selected and considering the contextual factors and their interdependencies in the experimental evaluation that leads to a specific technique selection. Furthermore, these factors can change over time, which creates an additional layer of complexity in the selection. 
In general, although the current literature provides a wide variety of ML techniques and algorithms, there is a lack of design approaches to support algorithm selection. 

In previous work, we have presented a preliminary investigation of potential variabilities in the ML model selection process analysis based on the CRISP-DM modeling phase \cite{tavares2021towards}. We have described initial feature models for the CRISP-DM modeling phase based on our initial understanding of the Scikit-Learn model selection heuristics. We have also developed a preliminary framework design for the CRISP-DM modeling phase. We have extended this preliminary investigation to focus on variabilities, feature diagrams, and constraints, based on the Scikit-Learn heuristics, that trigger adaptive reconfiguration, that is, changes in model selection due to changes in the variability factors \cite{tavares2022adaptive}. This previous work, accompanied by initial performance assessments using simple examples, indicates that modeling variabilities using feature diagrams and constraints was a promising research direction toward enhanced model selection methods.   
 
In this paper, we present a variability-aware ML algorithm selection approach that takes into account the commonalities and variations in the model selection process. In contrast with previous work, we present a design approach to support five phases: (i) identify and model variabilities and features; (ii) select a heuristic example and instantiate its feature diagram;(iii) design experimental case study; (iv) select model techniques; and (v) evaluate results. The proposed approach is comprehensive in the sense that it is structured through these five phases, shows general feature diagrams to support ML modeling technique selection and modeling assumptions, and applies these diagrams to represent ML algorithm selection based on Scikit-Learn. The approach also describes the design of an experimental case study, shows how suggested modeling techniques can be generated, and describes through a performance evaluation how the suggested techniques can be compared to techniques used in papers addressing ML modeling analysis in the literature. The design approach is beneficial in that it can suggest modeling techniques in a systematic way, as opposed to ad-hoc and informal, based on the factors that affect ML modeling selection and their interdependencies.    
The design approach involves identifying the variabilities in the ML algorithm selection phase and representing these variabilities through feature models. In addition, the approach also allows model selection feature models to be applied to represent existing procedures (e.g., heuristics) and uses these representations to guide the experiments that support a specific model technique selection. 
The applicability of the approach is illustrated by an experimental case study in which existing model selections presented in the literature are compared with selections suggested by the approach. 

Variability is the central concept of the proposed approach as we work on identifying and representing variabilities in ML model selection. This concept is defined as the ability of a product, an item, or a feature to change, evolve, or be customized \cite{galster2013variability, svahnberg2005taxonomy}. Variability-aware approaches have been proposed in several areas, including software engineering, databases and data warehouses \cite{valdezate2022ruva, Buhne2005, caplinskas2013variability, bouarar2015spl}. Designing and implementing a software approach to accommodate possible variations in factors that affect ML model (or algorithm) selection can lead to a highly configurable solution. The solution is configurable in the sense that the selection of specific variations (e.g., sample size, number of features, data type, prediction type) and their interdependencies can be sufficient to indicate which possible algorithm(s) can be applied in a concrete case. 

The feature model is one of the methods proposed in variability modeling analysis to formally represent the description of features and their constraints \cite{Kang2013}. Constraints in feature models denote conditions in which adaptations in variations occur. Feature models are visually represented by means of feature diagrams. This diagram is a tree-like structure used to represent features of a concept, where the root represents the main concept and the descendant or child nodes represent its features. Further, this model categorizes relationships between parent features and child (or sub-features). These relationships can be {\it mandatory} (i.e., a child feature must be selected), {\it optional} (i.e., a child feature is optional, that is, it can be selected or not selected), {\it or} (i.e., at least one of the child features must be selected), and {\it alternative} or {\it xor} (i.e.,  exactly one of the child features must be selected). For example, if the concept being modeled is a car, the body and the engine are mandatory features, music player and camera are optional features. Electric, petrol or gas can be alternative features, and monochromatic or polychromatic colour can be or features. In addition to parental relationships between features, cross-tree constraints are also allowed. These constraints or rules can specify, for instance, that a constraint can state that the selection of a feature can imply the selection of another feature. For example, a car needs to be electric to have an electric window opener, so the selection of a feature that supports an electric window opener requires the feature electric. In addition, a feature diagram can be instantiated as an application in the sense that a specific subset of features defined in the feature diagram that complies with the diagram relationships and constraints is chosen. For example, an application representing a specific car can have a subset of features such as car body, engine and music player and gas.

The work advances the state of the art in the development of methods to support the design and automation of ML algorithm selection in ML application development. The proposed approach can benefit both designers and practitioners, as it can lead to cost and time savings. In addition, it can make the selection more accessible and understandable to non-expert users. Further, this approach can be seen as a step towards the provision of a more explicit, adaptive, transparent, interpretable, and automated basis for model selection.

\subsection{Goal}
The goal of the proposed approach is to define a variability-aware approach to ML model selection in which the selection adapts to variations in the factors that affect ML model selection. The adaptation can be static or dynamic. In the static sense, different algorithms can be selected based on factors such as sample size, number of features, data type, prediction type, or non-functional factors such as performance (e.g., metrics such as accuracy, precision, recall, and F1-score) and ethics (e.g., fairness metrics).

Although the focus of this paper is on static, design-time adaptations, in the dynamic sense, data can change over time, affecting model selection and its outcomes and performance. In conventional ML modeling, the data sample and the problem are defined, and the selection of an algorithm may follow known heuristics. An illustrative dynamic scenario consists of situations in which the data characteristics or attributes change over time. For example, in the evaluation of the side effects of a new drug, the data is initially unlabeled, and the sample size is small. After a while, additional data is incorporated into the sample, and the data can become larger and labeled. In this case, the algorithm selection would have to be revisited to ensure model accuracy. 

\subsection{Research Questions}
This research relies on the following research questions:
\begin{itemize}
\item \textbf{RQ1}: How to identify and model the factors that capture the variability of ML model selection and their interdependencies based on feature diagrams?

\item \textbf{RQ2}: How can variability models be instantiated to capture specific model selection procedures(e.g., heuristics)?

\item \textbf{RQ3}: How to design an experimental case study to demonstrate the applicability of our approach based on specific performance metrics?
\end{itemize}

\subsection{Variability-Aware Method}
This paper presents an adaptive variability-aware method for ML model selection. Essentially, the proposed method involves:

\begin{enumerate}
    \item identify the factors that affect model selection proposed in the literature and represent these factors using a feature diagram and constraints (Section IV); 
    \item select a specific model selection procedure, that is, a model selection heuristics, and instantiate the general feature diagrams to represent these heuristics (Section IV); and 
    \item design an experimental case study that compares the techniques used in the experiment described in the literature with the techniques proposed by the approach using specific metrics (Section V).
\end{enumerate}

\subsection{Benefits}
The proposed approach can lead to an improved understanding of what factors influence model selection, how these factors explicitly affect selection, and how the adaptive factors can be represented and automated. This improved understanding can result in a project model selection process that is less implicit and more productive.
The proposed method also advances the state of the art by introducing an adaptive and interpretable process. Introducing adaptive processes provides support in dealing with the variations that occur in model selection. In addition, introducing an explainable process provides support for accountability, making clear the reasons why a method has been selected.      
Finally, the proposed adaptive method can ultimately constitute a foundation for the creation of novel software product lines to support the model selection process.

\subsection{Paper Structure}
The paper is structured as follows. Sections II and III describe the research background and related work, respectively. Section IV presents our variability-aware ML model selection approach. Section IV presents an experimental case study that illustrates the applicability of the approach. Section VI presents a discussion, and finally, Section VII presents conclusions and future work.   

\section{Background}
\subsection{ML Model Selection}
Modeling is one of the data analysis process phases, which consists of selecting and applying several modeling techniques and their algorithms to solve a problem until specific quality criteria are satisfied \cite{chapman2000crisp}. Among the plethora of model technique types and algorithms, ML-based data analysis is a widely adopted paradigm that has been applied in data science problems in several domains, including health, business, and smart cities \cite{francca2021overview}. Considering that different ML-based algorithms can be selected based on factors such as sample size, method category, and the types of data, selecting the appropriate algorithm is one of the most challenging steps in the context of the data model selection process. According to \cite{yao2018taking}, every aspect of ML-based analysis applications needs to be configured, indicating a need for new approaches and systems to automate the various phases of the data analysis process. 
Leenings et al. \cite{leenings2021photonai} presents an approach to accelerate machine learning model development that includes an estimator selector module. They argue that knowing a priori the optimal learning algorithm for a specific task is impossible, so their estimator selector method consists of the execution of several ML algorithms to identify the most efficient one. 

\subsection{Heuristics in ML Model Selection}
A few heuristics have been proposed to capture criteria for choosing particular machine learning methods. These heuristics can help automate this selection if they are appropriately captured. Although heuristics have been provided to guide the ML data analysis modeling process phases, such as the algorithm selection phase, these heuristics have not been used to capture the variability of this phase. From a software engineering perspective, designing and implementing a software approach to accommodate possible variations in factors that affect ML algorithm selection based on heuristics can lead to a configurable solution.

Some heuristics have been proposed to capture ways to select specific machine learning techniques. If captured appropriately, these heuristics can contribute to the automation of this selection. 
Scikit-Learn 1.0.1 \cite{pedregosa2011scikit} provides a flowchart with heuristics to guide users on how to select specific algorithms. Likewise, Microsoft \cite{microsoft-heuristic} currently released an informative sheet characterized as a starting point for algorithm selection that offers suggestions to users within their ML platform, Azure ML Studio. Accordingly, these suggestions are general rules-of-thumb and do not substitute deep knowledge of how to use ML algorithms. 

\subsection{Variability and Feature Modeling}
Variability is defined as the ability of a product, an item, or a feature to change, evolve, or be customized \cite{svahnberg2005taxonomy}. Identifying and representing variabilities constitute a basis for exploring opportunities for automation. Variability-aware approaches have been proposed in several areas, including software engineering, databases and data warehouses \cite{valdezate2022ruva, Buhne2005,caplinskas2013variability, bouarar2015spl}. 

Commonality and variability of a product can be captured in an abstract way using entities called features \cite{berger2015feature}. According to Kang et al. \cite{Kang2013}, a feature is “a prominent or distinctive and user-visible aspect, quality, or characteristic of a software system or systems.” Features are a key concept in variability modeling. A feature provides an abstract view of a variable and common requirements of a domain. Feature modeling is a technique widely adopted to model common and variable attributes \cite{Kang2013}. The feature model is one of the methods proposed in variability modeling analysis to formally represent the description of features and their constraints. Constraints in feature models can be seen to denote the conditions in which adaptations in variations occur. 

Feature modeling is a technique widely adopted to model common and variable attributes \cite{Kang2013}. Commonality and variability of a product can be captured in an abstract way using entities called features \cite{berger2015feature}. According to Kang et al.\cite{Kang2013}, a feature is “a prominent or distinctive and user-visible aspect, quality, or characteristic of a software system or systems.” Features are a key concept in variability modeling. A feature provides an abstract view of the variable and common requirements in a domain.

\section{Related Work}
Related work encompasses topics such as variability, automation and monitoring changes in the ML application lifecycle process and ML-based process reconfiguration.

Regarding variability, although variabiliy-aware approaches have been proposed in several domains \cite{valdezate2022ruva, Buhne2005, caplinskas2013variability, bouarar2015spl}, there is a lack of software design and implementation approaches to 
accommodate possible variations in factors that affect ML model selection and support highly configurable solutions. Current approaches can not support model selection based on specific variations such as those related to sample size, number of features, data type, prediction type, and non-functional requirements such as performance and fairness
 \cite{galster2013variability, svahnberg2005taxonomy}.

Variability is the central concept of the proposed approach as we work on identifying and representing variabilities in ML model selection. This concept is defined as the ability of a product, an item, or a feature to change, evolve, or be customized \cite{galster2013variability, svahnberg2005taxonomy}. Variability-aware approaches have been proposed in several areas, including software engineering, databases and data warehouses \cite{valdezate2022ruva, Buhne2005, caplinskas2013variability, bouarar2015spl}.

Regarding automation, the acronym MLOps (Machine Learning Operations) refers to an approach to automating ML lifecycle processes aiming to accelerate the deployment of models in production, applying principles of continuous integration, delivery, and model retraining \cite{symeonidis2022mlops}.

MLOps pipeline comprises a cycle of tasks such as data preparation, model creation, training, evaluation, deployment, and monitoring, which are grouped into three main basic procedures, namely data manipulation, model creation, and deployment.

Regarding monitoring, Martinez \cite{rivera_2010} proposes a model monitoring approach that considers the factors that can affect the performance of the model, such as a change in data context (e.g., culture, location or time). The idea is to reconfigure the ML model based on the system performance dynamically. However, performance decrease is a critical problem when application data constantly changes over time, such as those related to healthcare scenarios, environmental monitoring, and air traffic control. 

Regarding reconfiguration, Nascimento et al. \cite{nascimento2021context} presents an example of a reconfigured neural network based on the application context. The authors describe situations in which it is necessary to modify the neural network architecture itself (e.g., the number of layers) and not only to retrain it. 

\begin{figure*}[!ht]
	\centering
    \includegraphics[width=\linewidth]{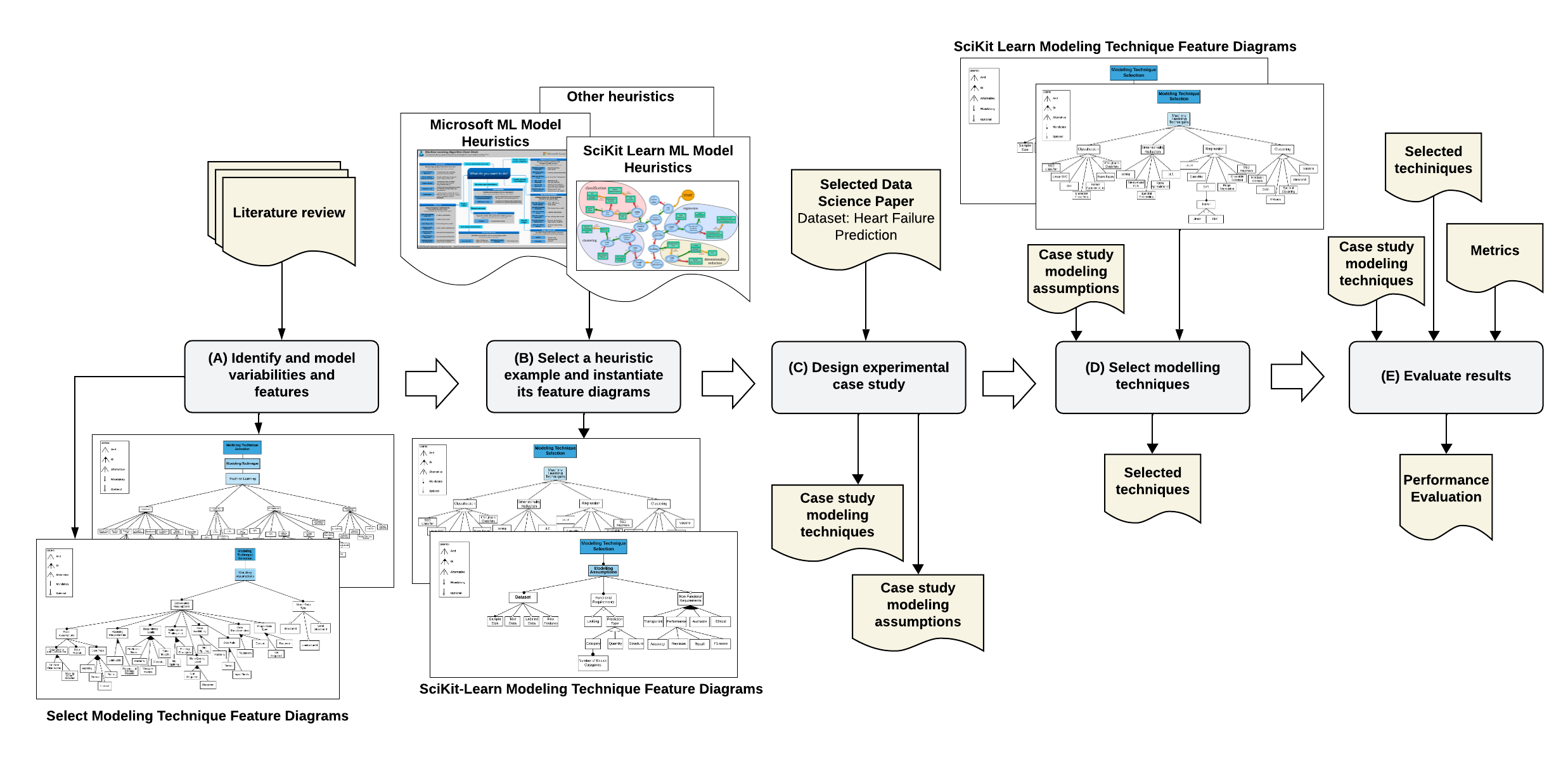}
	\centering
	\caption{Variability-aware ML algorithm selection approach.}
	\label{fig:Approach}
\end{figure*}

\section{Variability-Aware ML Algorithm Selection Approach} \label{Approach}
Figure \ref{fig:Approach} represents the phases of the proposed approach. The purpose of the figure is to visually represent the approach overview and the flow between each of the phases of the approach and the input and output elements of each phase, which are presented as larger figures in each subsection that describes the phases of the approach. These five phases involve: \\
\textbf{(A) Identify and model variabilities and features.} In this phase, based on the literature, we identify the factors that capture variabilities that affect ML model selection and represent these variabilities in general using feature diagrams. \\
\textbf{(B) Select a heuristic example and instantiate its feature diagram.} This phase aims at selecting an existing ML model selection heuristics, capturing its variabilities and constraints, and instantiating the feature diagrams derived in the previous phase based on the specific heuristics variabilities and potential extensions. \\
\textbf{(C) Design experimental case study.} In this phase, we design a use case by selecting an experiment described in the literature in which an ML algorithm is selected, using a heuristic to select an ML modeling algorithm to demonstrate the application of our approach. \\
\textbf{(D) Select modeling techniques.} This phase uses the modeling assumptions captured in the previous phase and the Scikit-Learn-based feature diagrams derived in phase (B) to generate possible configuration instances related to the experiment. \\
\textbf{(E) Evaluate results.} This phase aims to evaluate performance using specific metrics and analyze the results, comparing the performance of the models suggested by the heuristics with the models used in a specific experiment described in the selected paper. The chosen model or multiple models are those that have the best performance. //

The following subsections describe these phases: Phase A addresses research question RQ1, Phase B addresses RQ2, and Phases C, D, and E address RQ3.

\subsection{Identify and Model Variabilities and Features}
To build high-quality models, selecting the most appropriate ML algorithm is critical and depends on several key factors, such as data type and size, the type of problem to be solved, and the expected performance according to different metrics. In this phase, we identify and model these factors, which capture the variability of ML model selection and their interdependencies using feature diagrams. The identification of these factors relied on a literature review \cite{chapman2000crisp, project2017edison, KashyapPatanjali2023MLAa}, and two feature diagrams were developed. The review uncovered variabilities related to modeling techniques, dataset characteristics, functional requirements, and non-functional requirements related to quality attributes such as performance. 

\begin{figure*}[!ht]
	\centering
	\includegraphics[width=\linewidth]{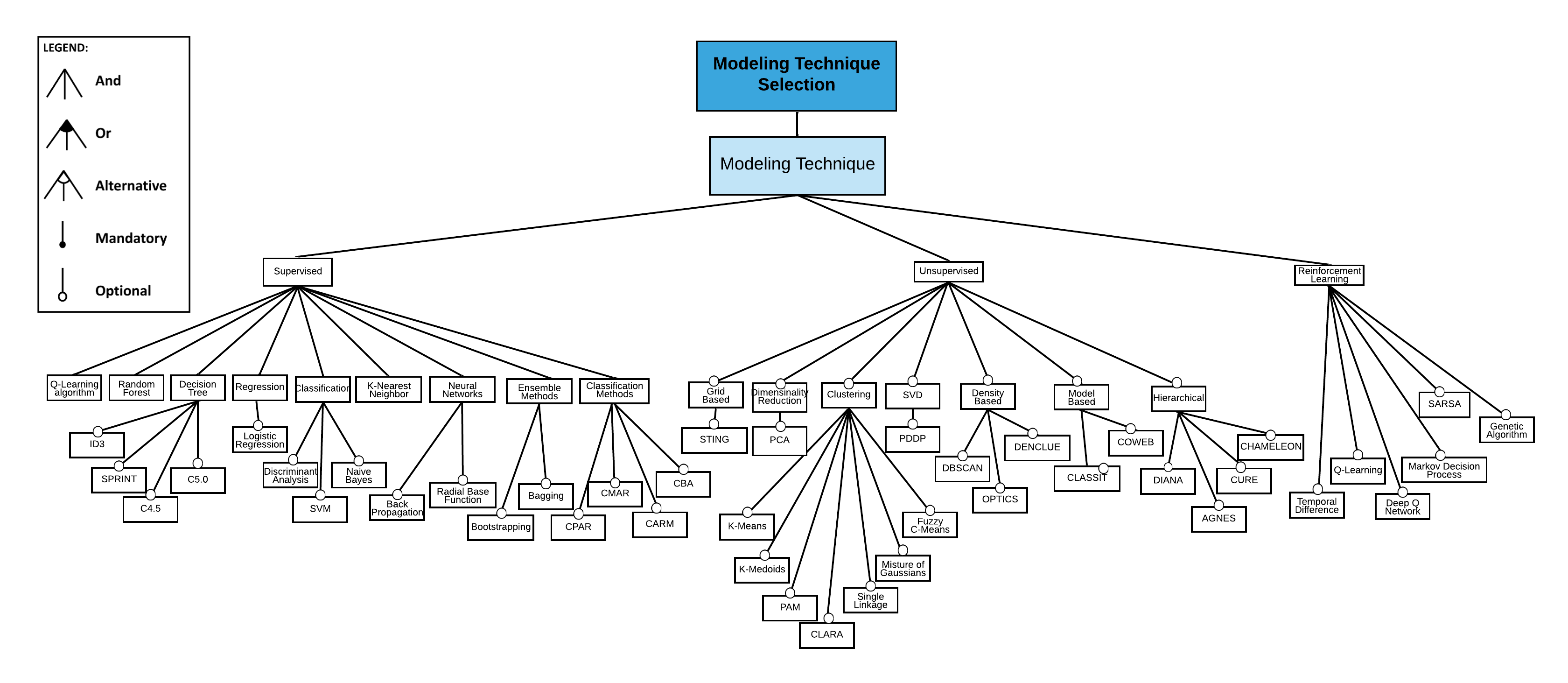}
	\centering
	\caption{Feature diagram for ML modeling technique selection.}
	\label{fig:Techniques}
\end{figure*}

\begin{figure*}[!ht]
	\centering
	\includegraphics[width=\linewidth]{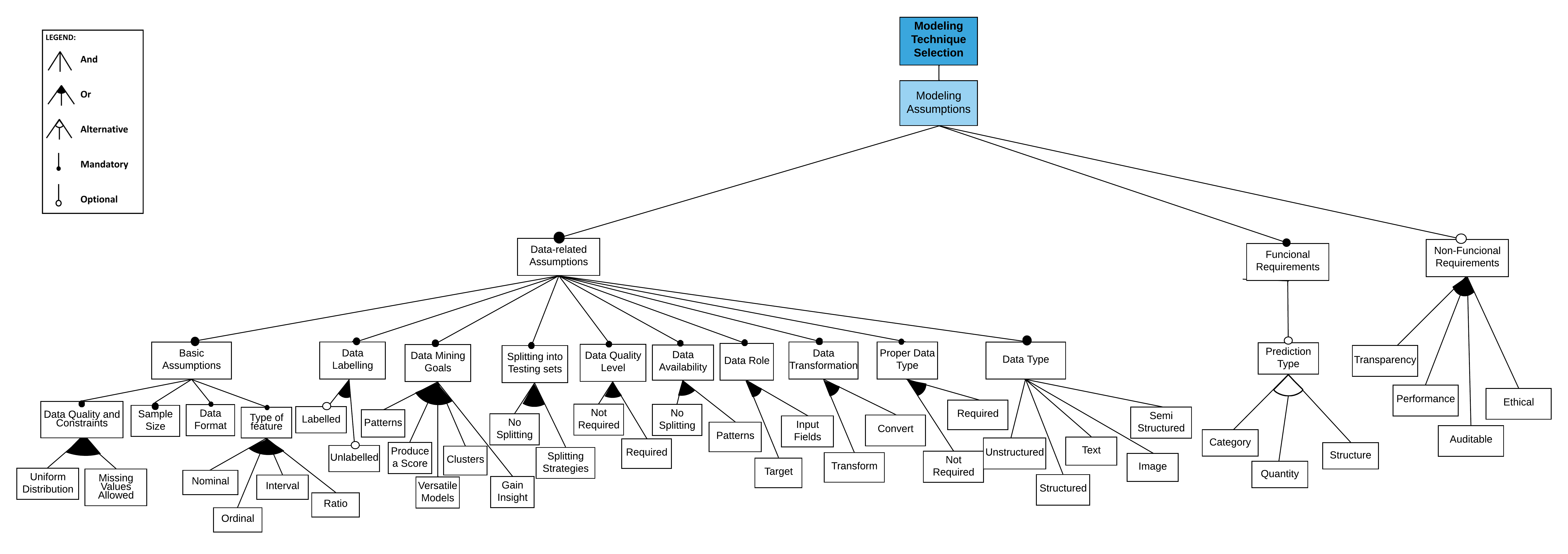}
	\caption{Feature diagram for ML modeling assumptions.}
	\label{fig:Assumptions}
\end{figure*}

Figures \ref{fig:Techniques} and \ref{fig:Assumptions} illustrate the feature diagrams that we designed to capture various types of variability factors involved in the process of selecting an ML algorithm. 

Figure \ref{fig:Techniques} shows the feature diagram for ML modeling technique selection. This diagram includes techniques involving ML supervised, unsupervised, and reinforcement learning methods.  

Figure \ref{fig:Assumptions} shows the feature diagram for ML modeling assumptions. In this diagram, the mandatory Assumptions feature Modeling consists of a group of alternative features which specify assumptions about data according to the modeling technique selected \cite{chapman2000crisp}. The activities for this step involve \cite{chapman2000crisp}: defining any built-in assumptions made by the technique about the data (e.g., quality, format, and distribution) and ensuring that the appropriate model would also need to consider the availability of data types for mining, the data mining goals, and the specific modeling requirements. These elements are optional and might occur depending on the technique selected. For example, applications with the data mining goal of predicting credit risk demand transparent, auditable, and explainable models \cite{ariza2020explainability}. In this case, decision trees are recommended because of their interpretability  \cite{ariza2020explainability}. In the medical field, applications also require higher levels of safety and explainability. Thus, logistic regression has been encouraged to develop explainable clinical predictive models, even when modern ML models outperform them \cite{zihni2020opening}. Based on the notion of optional features, the feature diagram depicts or-group features, namely Data-related Assumptions, Functional Requirements, and Non-Functional Requirements.

\subsection{ Select Heuristics Example and Instantiate Feature Diagram}

This phase describes how variability-aware models developed in the previous phase can be instantiated to capture specific model selection heuristics.  
Some approaches for ML model selection based on heuristics have been defined in the literature, including Scikit-Learn and Microsoft ML Designer \cite{pedregosa2011scikit, microsoft-heuristic}.

Scikit-Learn is an easy-to-use Python package focused on providing non-ML-specialists with a wide range of state-of-the-art ML algorithms to solve supervised and unsupervised problems\cite{pedregosa2011scikit}. 

\begin{figure*}[!ht]
	 \centering
 	\includegraphics[width=\linewidth]{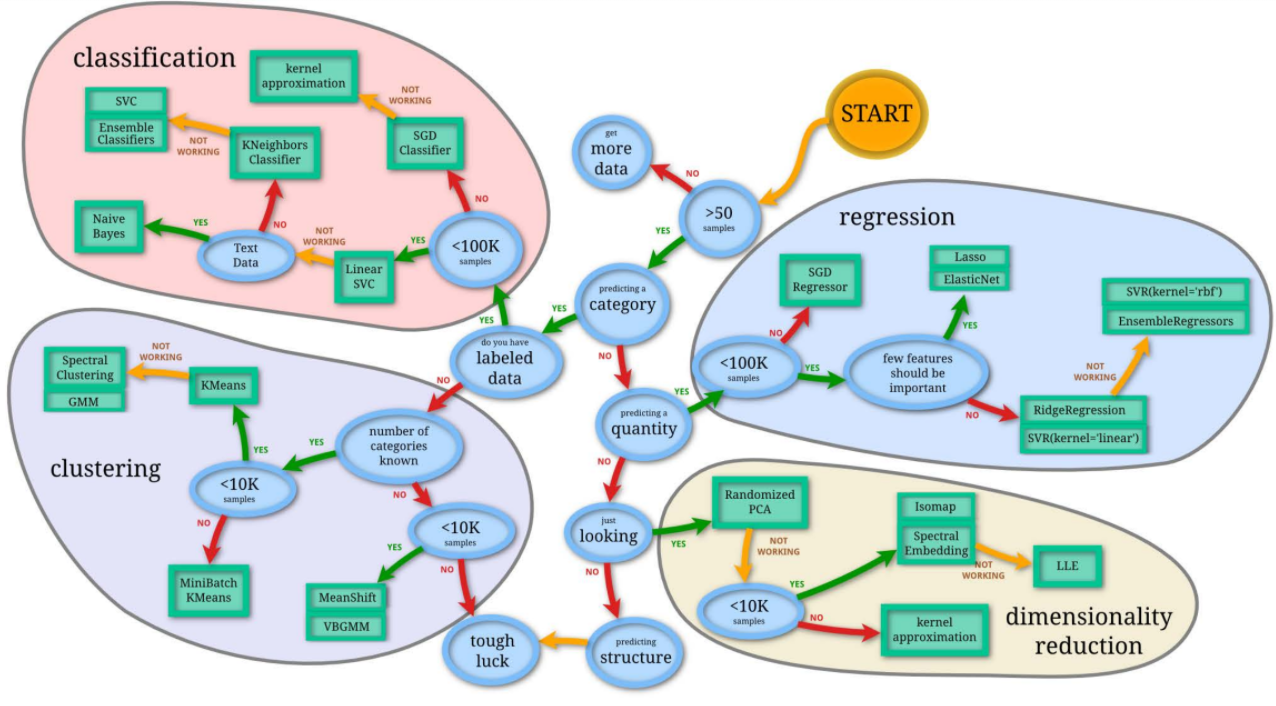}
	\centering
	\caption{Scikit-Learn flowchart for selecting a machine learning algorithm.}
	\label{fig:Scikit}
\end{figure*}

To understand the Scikit-Learn heuristics, we focused specifically on the flowchart provided in Scikit-Learn 1.0.1 \cite{pedregosa2011scikit}, shown in Figure \ref{fig:Scikit}. This flowchart provides guidance to support users on how to select specific algorithms. The Scikit-Learn flowchart establishes a set of rules to find the algorithm that is better suited for specific problems and types of data and also indicates some choices that can result in errors. Based on these rules, we captured a set of constraints (i.e., restrictions on the possible selections of features) for the feature diagram. These constraints capture the features that are required and excluded so that a specific algorithm is selected. For example, if the sample size is equal to or greater than 50, the prediction type could be a category. Then, following the flowchart, if the data is labelled, and the sample size is less than 100,000, then a linear SVC algorithm is recommended. In the case SVC is `not working,' and the data is textual, the next recommended algorithm is the Naive Bayes. 

\begin{figure*}[!ht]
	\centering
	\includegraphics[width=\linewidth]{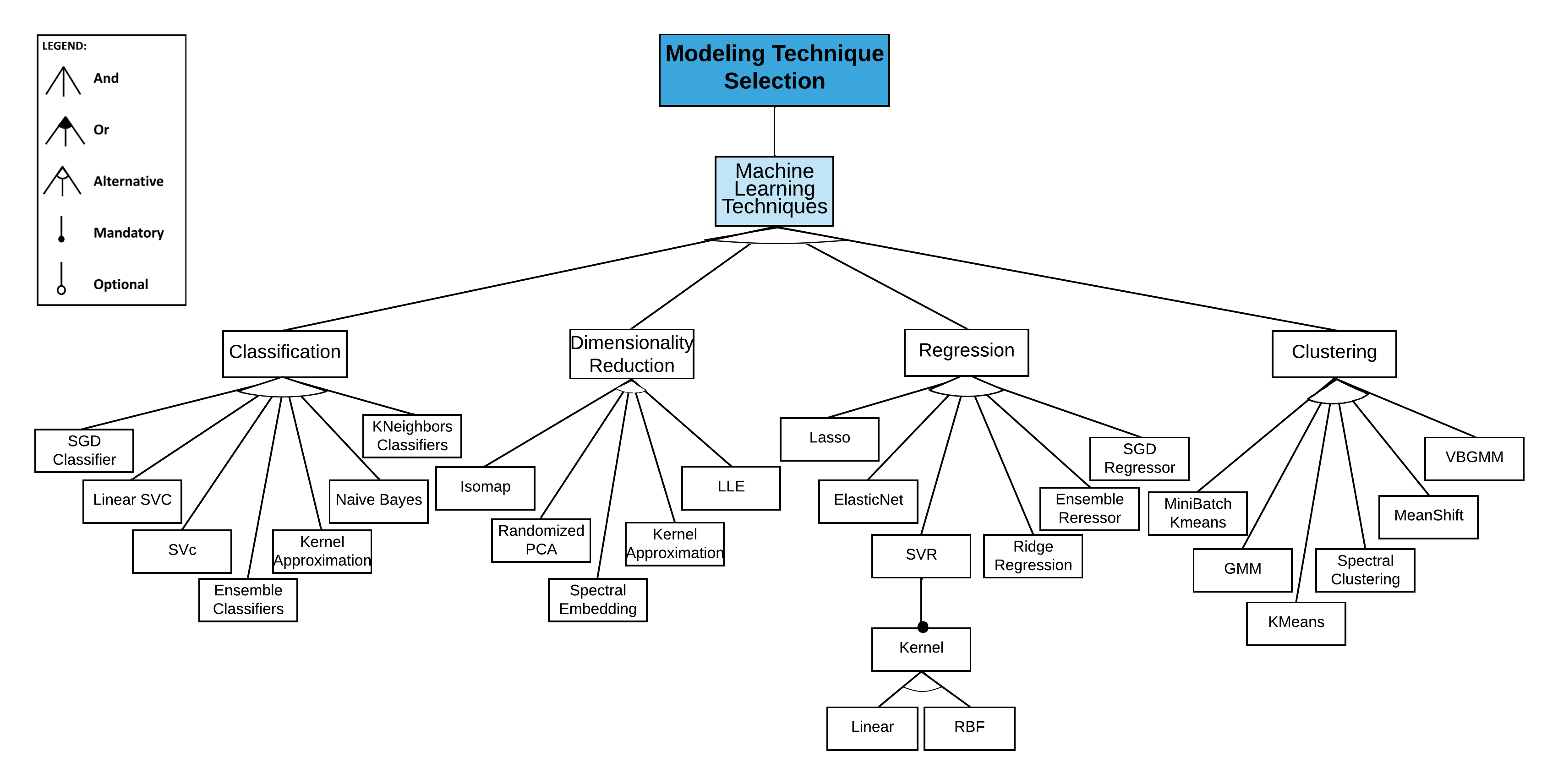}
	\centering
	\caption{Feature diagram of the instance of a ML algorithm selection from Scikit-Learn.}
	\label{fig:Scikit_Techniques}
\end{figure*}

\begin{figure*}[!ht]
	\centering
	\includegraphics[width=\linewidth]{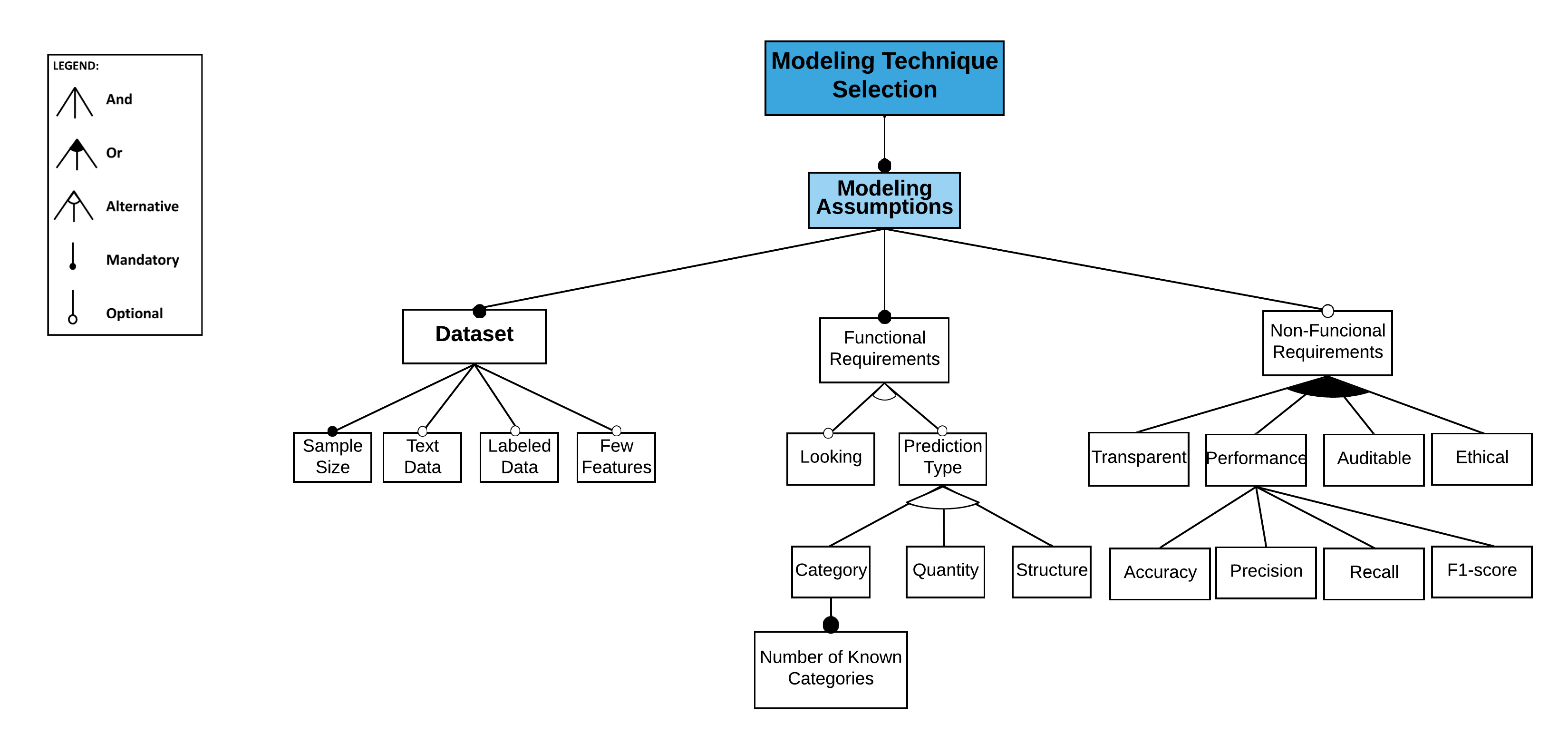}
	\centering
	\caption{Feature diagram of the instance of a ML algorithm assumptions from Scikit-Learn.}
	\label{fig:Scikit_Assumptions}
\end{figure*}

The Scikit-Learn flowchart establishes a set of rules to find the algorithm that is better suited for specific problems and types of data, also illustrating some choices that can result in errors. Based on the Scikit-Learn heuristics, we applied the feature modeling technique to capture variability factors and their interdependencies as constraints, and represented them as two feature model diagrams. 

The model selection diagram, presented in Figure \ref{fig:Scikit_Techniques}, depicts the feature diagram representing Scikit-Learn modeling techniques, which include Classification, Dimensionality Reduction, Regression, and Clustering. For example, classification techniques include Linear SVC, SGD classifier, SVC, Naive Bayes, and KNeighbors Classifier.

The second diagram, presented in Figure \ref{fig:Scikit_Assumptions}, shows the feature diagram considering Scikit-Learn modeling assumptions, which include dataset requirements (e.g., sample size), functional requirements (e.g., prediction type), and non-functional requirements such as performance and ethical considerations (e.g., fairness). Performance metrics include accuracy, precision, recall, and F1-score. 

The Scikit-Learn flowchart represents a 'not working' criteria as a factor when selecting another algorithm type. However, the Scikit-Learn flowchart does not specify the criteria for defining a 'not working' model. To address this gap, we extended the instantiated Scikit-Learn feature diagram to provide alternatives for representing the 'not working' that considers that a model is not working when it does not satisfy non-functional constraints such as performance or other metrics.

These two feature diagrams presented in Figures \ref{fig:Scikit_Techniques} and \ref{fig:Scikit_Assumptions}, respectively, correspond to instances of the feature diagrams for Modeling Techniques and Modeling Assumptions presented in Figures \ref{fig:Techniques} and \ref{fig:Assumptions}. 

Further, the Scikit-Learn heuristics establish a set of rules or constraints that govern the relationships between features beyond the basic hierarchical structures. These rules guide the selection of the algorithm that is better suited for specific model selection problems. 

Below, we present these Scikit-Learn constraints, which are listed according to prediction type selection, modeling technique category, regression methods, classification methods, and clustering methods. These constraints are represented in logic, using logical connectives such as conjunction (and), $\wedge$; disjunction (or), $\vee$; implication (if...then), $\Rightarrow$; negation (not), $\neg$; and equivalence (if and only if), $\Leftrightarrow$.

C1 - Constraints for selecting prediction type:

\begin{quote}
{\footnotesize   
\(1. Category \Leftrightarrow \neg(Quantity \wedge  Structure)\)
}
\end{quote}

Following the Scikit-Learn heuristics (Figure 4), this expression means that predicting a Category is equivalent to not predicting a Quantity and a Structure. Note that, according to these heuristics, to use any modeling technique option from the Scikit-Learn library, it is necessary to have a dataset with more than 50 samples.

\hspace*{0.1cm}

C2 - Constraints for selecting modeling technique category:

\begin{quote}
{\footnotesize   
\(1. Samplesize > 50 \wedge Predictiontype \wedge Quantity \Leftrightarrow Regression\)
}
\end{quote}

Following the Scikit-Learn heuristics, this expression means that having a Samplesize greater than 50 and conducting some Predictiontype that involves a Quantity is equivalent to conducting a Regression. Additional constraints for selecting modeling technique category are: 

\begin{quote}
{\footnotesize 
\(2. Samplesize > 50 \wedge Predictiontype \wedge Category \wedge LabeledData \Leftrightarrow Classification\)

\(3. Samplesize > 50 \wedge Predictiontype \wedge Category \wedge \neg LabeledData \Leftrightarrow Clustering\)

\(4. Samplesize > 50 \wedge \neg Predictiontype  \Leftrightarrow DimensionalityReduction\)
}
\end{quote}

\hspace*{0.1cm}

C3 - Constraints for selecting regression methods:
\begin{quote}
{\footnotesize   
\(1. Regression \wedge Samplesize < 100K \wedge Fewfeatures \Rightarrow Lasso \lor ElasticNet \)
}
\end{quote}

This expression means that if a regression is conducted and Samplesize is less than 100K and there are few features, then the Lasso or the ElasticNet modeling technique should be selected. Additional constraints for selecting regression methods are: 

\begin{quote}
{\footnotesize 
\(2. Regression \wedge Samplesize < 100K \wedge \neg Fewfeatures \Rightarrow RidgeRegression \lor (SVR \wedge linear) \)

\(3. RidgeRegression \lor (SVR \wedge linear) \wedge \neg notWorking \Rightarrow ((SVR \wedge rbf) \lor EnsembleRegressors)\)

\(4. Regression \wedge Samplesize >= 100K \Rightarrow SGDRegressor \)
}
\end{quote}

\hspace*{0.1cm}

C4 - Constraints for selecting dimensionality reduction methods:
\begin{quote}
{\footnotesize   
\(1. DimensionalityReduction \Rightarrow RandomizedPCA   \) 
}
\end{quote}

This expression means that if DimensionalityReduction is conducted then the RandomizedPCA modeling technique should be selected. Additional constraints for selecting dimensionality reduction methods are:

\begin{quote}
{\footnotesize 
\(2. RandomizedPCA \wedge \neg notWorking \wedge Samplesize < 10K \Rightarrow Isomap \lor SpectralEmbedding   \) 

\(3. RandomizedPCA \wedge (Isomap \lor SpectralEmbedding) \wedge \neg notWorking \wedge Samplesize < 10K \Rightarrow  LLE  \) 

\(4. RandomizedPCA \wedge \neg notWorking \wedge Samplesize >= 10K \Rightarrow kernelApproximation   \) 
}
\end{quote}


C5 - Constraints for selecting classification methods:
\begin{quote}
{\footnotesize   
\(1. Classification \wedge Samplesize < 100K \Rightarrow  LinearSVC\)
}
\end{quote}

This expression means that if Classification is conducted and Samplesize is less than 100K then the LinearSVC modeling technique should be selected. Additional constraints for selecting classification methods are:

\begin{quote}
{\footnotesize 
\(2. LinearSVC \wedge \neg notWorking \wedge  Textdata \Rightarrow  NaiveBayes\)

\(3. LinearSVC \wedge \neg notWorking \wedge  \neg Textdata \Rightarrow  KNeighbors Classifier\)

\(4. LinearSVC \wedge KNeighborsClassifier \wedge \neg notWorking \wedge \neg Textdata \Rightarrow  (SVC \lor EnsembleClassifiers)\)

\(5. Classification \wedge Samplesize >= 100K \Rightarrow  SGDClassifier\)

\(6. SGDClassifier \wedge \neg notWorking \wedge Samplesize >= 100K \Rightarrow  kernelApproximation\)
}
\end{quote}

C6 - Constraints for selecting clustering methods:
\begin{quote}
{\footnotesize   
\(1. Clustering \wedge Knowncategories \wedge Samplesize >= 10K \Rightarrow MiniBatchKMeans\)

\(2. Clustering \wedge Knowncategories \wedge Samplesize < 10K \Rightarrow KMeans\)

\(3. KMeans \wedge \neg notWorking \Rightarrow SpectralClustering \lor GMM\)

\(4. Clustering \wedge \neg Knowncategories \Rightarrow  Samplesize < 10K \)

\(5. Clustering \wedge \neg Knowncategories \wedge Samplesize < 10K  \Rightarrow MeanShift \lor VBGMM \)  
}
\end{quote}

\subsection{Design Experimental Case Study}
The purpose of this phase is to design an experimental case study. The study is based on a selected paper that uses one or more ML modeling techniques and provides experiments and results. The output of this phase is the specification of the modeling techniques and assumptions based on the paper.   

\subsection{Selecting modeling techniques}
In this phase, we generate selected modeling techniques using as input the feature diagrams captured from the heuristics used in the example application and the modeling assumptions identified in the experimental case study. 

\subsection{Evaluate Results}
The purpose of this phase is to compare the case study modeling techniques with the techniques selected by our approach by undertaking a performance evaluation using specific metrics and analyze the results.

\section{Case Study}
In this section, we present a case study to show how the approach can be used for ML modeling technique selection. The case study involves and experimental design illustrated in Figure \ref{fig:Approach} as phases C, D and E, which are briefly described in the previous section.

\subsection*{Phase C - Design Experimental Case Study}
The experimental design requires the selection of an experiment described in a paper that uses one or more ML modeling techniques and provides explicit experimental results. We have identified a study conducted by Chicco and Jurman \cite{chicco2020machine} in which the application purpose is to predict the survival of patients with heart failure based on clinical information using several ML modeling techniques. Another study conducted by Leenings et al. \cite{leenings2021photonai} also aims at predicting patient survival and relies on the same dataset used by Chicco and Jurman \cite{chicco2020machine}. The study proposed by Leenings et al. outlines an approach for automating the selection of machine learning algorithms and also employs the Scikit-Learn library to test their solution.

For this experiment, we used the Scikit-Learn feature diagrams provided in phase B of the approach (Figure \ref{fig:Approach}). 

\subsubsection{Dataset description}
The application dataset consists of data from 299 patients with heart failure. This dataset comprises 13 clinical features that can be used to predict mortality by heart failure, such as age, sex, if the patient has diabetes, and serum creatinine level in the blood. This dataset contains an additional feature used as the target in the classification study, called 'death\_event', that is a binary measure that indicates if the patient died or survived before the end of the follow-up period, which was 130 days on average. The dataset is imbalanced because the number of patients who survived (death event = 0) is 203, while the number of those who died (death event = 1) is 96. In statistical terms, there are 32.11\% positives and 67.89\% negatives.  

\subsubsection{Experiment modeling techniques and assumptions} 

The classification techniques used by Chicco and Jurman \cite{chicco2020machine} include Linear Regression, Random Forest, Decision Tree, Artificial Neural Network (i.e., Perceptron), Support Vector Machines (Linear and Gaussian Radial Kernel), k-Nearest Neighbors, Naive Bayes, and Gradient Boosting. Leenings et al. \cite{leenings2021photonai} tested Random Forest, Gradient Boosting, and a Support Vector Machine for the same dataset. 

The assumptions for this case study include: prediction type, data type, sample size, number of features, and labeled data. In our example, the prediction type involves predicting a category, whether the patient will survive or not. The data types of the features used by the model are not text-based (e.g., age, sex, or level of serum creatinine in the blood). The sample size, which is the number of patients, is 299. The number of features is 13. The example involves binary labeled data, based on the feature called 'death\_event,' which is a binary feature indicating if the patient has died or survived. In terms of non-functional requirements, the example evaluates the performance of the model according to the models' metrics, such as accuracy and F1-score.

\subsection*{Phase D - Select modeling techniques}

The purpose of this phase is to select the most suitable modeling technique to solve the experimental case study, which is based on the Scikit-Learn heuristics. 
These heuristics are modeled by the feature diagrams (see Figures \ref{fig:Scikit_Techniques} and \ref{fig:Scikit_Assumptions}) and constraints described in Phase B (Figure \ref{fig:Approach}). 
Following Scikit-Learn heuristics, the dataset attributes and the purpose of the application the most appropriate algorithm is selected. 

In this application, the dataset contains 299 samples (\textit{Sample Size}). Our experiment application involves the \textit{Prediction Type} of a \textit{Category} related to mortality by heart failure, which could lead to \textit{Classification} or \textit{Clustering}. The dataset has an attribute that indicates the patient died or survived - 'death\_event'-  suggesting the data is labelled (\textit{Labeled Data}). The requirements indicate \textit{Classification} as the ML algorithm to solve our experimental problem.  

After identifying the ML category type, the next step is to identify one of the algorithm options from this category as the best fit. According to Scikit-Learn, two requirements define the most appropriate technique for the classification category: the \textit{Sample Size} and if it is \textit{Text Data}. Following the Scikit-Learn heuristics (Figure \ref{fig:Scikit}), our dataset satisfies the criterion of having less than 100K samples, indicating that \textit{Linear SVC} could be a suitable technique to solve the problem. In this case, if it is not working \textit{(not working)} (see the classification bubble in Figure \ref{fig:Scikit})
and the data is not text \textit{(Text Data)}, it could lead to \textit{KNeighbors Classifier}, and if this one is not working, it would lead to \textit{SVC} or \textit{Enseble Classifiers}.

This step is also supported by constraints extracted from the Scikit-Learn flowchart and described in Phase B (Figure \ref{fig:Approach}). The constraints that guided the technique selection are:

C1 - Constraints for selecting prediction type:
\begin{quote}
{\footnotesize   
\(1. Category \Leftrightarrow \neg(Quantity \wedge  Structure)\)
}
\end{quote}

C2 - Constraints for selecting modeling technique category:
\begin{quote}
{\footnotesize   
\(2. Samplesize > 50 \wedge Predictiontype \wedge Category \wedge LabeledData \Leftrightarrow Classification\)
}
\end{quote}

C5 - Constraints for selecting classification methods:
\begin{quote}
{\footnotesize   
\(1. Classification \wedge Samplesize < 100K \Rightarrow  LinearSVC\)

\(3. LinearSVC \wedge \neg notWorking \wedge  \neg Textdata \Rightarrow  KNeighbors Classifier\)

\(4. LinearSVC \wedge KNeighborsClassifier \wedge \neg notWorking \wedge \neg Textdata \Rightarrow  (SVC \lor EnsembleClassifiers)\)
}
\end{quote}

Therefore, according to the constraints above, our approach suggests the following techniques: Linear SVC, KNeighbors Classifier, SVC, and Ensemble Classifiers. 

\subsection*{Phase E - Evaluate Selected Techniques}

In this section, we compare the case study modeling techniques with the techniques selected by our approach by undertaking a performance evaluation using specific metrics and analyzing the results. We evaluated the performance of the techniques suggested by our approach, assessing specific metrics. We compared our results and performance metrics with those found in other studies that used the same example experiment \cite{chicco2020machine, leenings2021photonai}. 
In summary, after comparing the performance of the models suggested by the heuristics with the models used in a specific experiment described in a selected paper, if the performance of one or multiple suggested models is greater than the performance of those used in the specific experiment, then this one model or multiple models can be adopted.

Based on the Scikit-Learn feature diagrams and constraints, our approach identified this application as a classification problem, and the suggested algorithms, according to the previous phase, are LinearSVC, KNeighborsClassifier, SVC, and EnsembleClassifiers. The feature diagram indicates that the best technique for solving this type of problem is Linear SVC, and as such, it is not necessary to pre-select other algorithms and run each one to evaluate the best-fit model. According to the Scikit-Learn heuristics, the method `not Working' is used in case an algorithm does not fit to solve that problem, indicating that one should move the next algorithm in the recommendation queue. For example, the next classification algorithm in case the data is not textual is KNeighbor Classifier.

Thus, we initiated our experiment by running the LinearSVC algorithm using the entire set of clinical features in the dataset. For the data splitting method, we also applied the same split strategy employed by Leenings et al. \cite{leenings2021photonai}. We split the dataset into 80\% (239 randomly selected patients) for the training set and 20\% for the test set (the remaining 60 patients). 
We also used a stratified split, which is a recommended approach for imbalanced datasets. 
To automate the evaluation of all model parameter combinations via cross-validation, we employed the GridSearchCV optimizer from Scikit-Learn.

To measure model performance, we calculated the F1-score, Matthews correlation coefficient, balanced accuracy, sensitivity, and specificity. We assumed the F1-score below a specific value represented the "not working'' criteria. The F1-score goal was set as 0.77, taking into consideration the baseline of 0.76, which was the highest F1-score found in Leening et al.'s work \cite{leenings2021photonai}. The F1-score found in Leening's \cite{leenings2021photonai} outperformed the one found in Chicco and Jurman \cite{chicco2020machine}. 

This application is a classification problem, and the most recommended classification algorithm for this dataset is the Linear SVC. 
According to our evaluation, the Linear SVC algorithm outperforms the results presented in \cite{leenings2021photonai}. Their final results were using the Random Forest algorithm, achieving the following results: F1=0.746, mattews corr=0.619, BACC=0.813, sens=0.813, and spec=0.823. 

\begin{lstlisting}[language=Octave]
Category: Classification
Selected Method: LinearSVC
Balanced Accuracy (BACC):  0.8478818998716302
Matthews correlation coefficient (mattews corr):  0.6716258743858474
Accuracy :  0.85
Sensitivity :  0.8536585365853658
Specificity :  0.8421052631578947
f1 score:  0.7804878048780488
\end{lstlisting}

Compared with findings in both works of Leenings et al. \cite{leenings2021photonai} and Chicco and Jurman \cite{chicco2020machine}, the model selected by our approach outperformed their results as presented in Table \ref{tab_performance}.

Leenings et al. \cite{leenings2021photonai} executed the Random Forest classifier, gradient boosting, and the support vector classification (SVC) with linear kernel to this dataset in order to identify the one that provides the best F1 score. After executing these algorithms, the authors identified that the Random Forest Classifier presented the best performance. The original paper \cite{chicco2020machine} also proposes the use of Random Forest. However, as Chicco and Jurman \cite{chicco2020machine} do not use an approach to deal with the dataset imbalance, they achieved a lower F1 score of 0.547.

Although the SVC with linear kernel algorithm is similar to the Linear SVC algorithm, which is the one that we used in our tests, they differ in terms of flexibility in the choice of penalties and loss functions \cite{pedregosa2011scikit}. 
The two other algorithms tested by Leenings et al. \cite{leenings2021photonai}, Random Forest Classifier and Gradient Boosting, are examples of Ensemble Classifiers. According to the Scikit-Learn flowchart, LinearSVC, KNeighborsClassifier, and SVC are more highly recommended for this dataset than Ensemble Classifiers. 

Leenings et al. \cite{leenings2021photonai} also tested the LassoFeatureSelection for automatically selecting features from the dataset, and they observed a decrease in the system performance. According to the Scikit-Learn constraints, the Lasso algorithm is recommended as a regression algorithm. 

\begin{table}
\caption{Performance metrics for the final model.}
\label{tab_performance}
\setlength{\tabcolsep}{1pt}
\begin{tabular}{l|l|c|c|c|c|c}
\hline
\textbf{Experimental}& \textbf{Final} & \textbf{F1-Score} & \textbf{MCC} & \textbf{BACC} & \textbf{Sens} & \textbf{Spec} \\
\textbf{Study}& \textbf{Estimator} &  &  &  &  &  \\\hline

Our Approach & LinearSVC & 0.780 & 0.672 & 0.848 & 0.854 & 0.842  \\\hline
Chicco et al.\cite{chicco2020machine}& Logistic \break{Regression} & 0.714 & 0.607 & 0.818 & 0.780 & 0.856 \\\hline
Leenings et al.\cite{leenings2021photonai}& Random Forest  & 0.746 & 0.619 & 0.813 & 0.813 & 0.823 \\\hline

\multicolumn{7}{p{235pt}}{Notes: MCC= Matthews correlation coefficient, BACC= balanced accuracy, Sens= sensitivity, Spec= specificity}

\end{tabular}
\end{table}

\section{Discussion}

\subsection{Addressed Gaps}

The proposed approach addresses several relevant gaps in ML modeling selection. First, this approach provides a more explicit and formal basis for the selection. There are implicit assumptions about the data that are used in algorithm selection that, beyond sample size and data type, include the presence of missing values and missing categorical values. The violation of these assumptions can result in the generation of wrong predictions  \cite{schelter2018deequ}. Second, the approach allows model selection procedures, such as heuristics, to be specified, which capture expert knowledge and constitute a rich design space that can inform variability models to help further the design and automation of ML algorithm selection. Finally, the proposed approach captures interactions among various abstractions that impact algorithm selection (e.g., dataset, prediction type, and outcomes) as these elements change over time. For example, the data sample size may increase and result in the selection of different algorithms, the prediction category may change as additional quantitative data becomes available, and new important features are discovered. They can be added to the dataset, or changes in outcomes (e.g., accuracy) may result in changes in the models (e.g., the drift problem \cite{svahnberg2005taxonomy, symeonidis2022mlops}. In more dynamic settings, according to Hummer et al. \cite{hummer2019modelops}, while in the classical application lifecycle, new builds are triggered by code base changes, in the AI application lifecycle, new builds could be triggered by data or code changes, which may activate a re-training process.

\subsection{Threats to Validity}
Threats to validity include the selection of an experiment in which several models were selected and evaluated in the literature. Although a different selection could lead to different results, the purpose of the experimental study is to evaluate the approach's capability to recommend models that, in some cases, can perform better than the models used in papers published in the literature. 

There are also threats related to the heuristics adopted in the study. These heuristics capture valuable expertise in model selection, and some of these heuristics have been used extensively by practitioners. However, these heuristics have not been formally evaluated in terms of the parameters they use and their correctness. Our purpose, in contrast, is not to evaluate the heuristics \textit{per se} but to introduce an approach that can be used to represent specific model selection procedures (e.g., heuristics), based on the factors that affect their variability and their interdependencies using feature diagrams and be utilized as a method for comparing how the models selected in specific experiments published in the literature fair in comparison with the modes suggested by the heuristics. The approach could be extended to compare the selection of different model selection procedures and different experiments to be conducted or provided in the literature.

\subsection{Dynamic Adaptations in ML Model Selection}
Although the focus of this paper is on design-time variabilities, we note that the proposed approach can also be applied to cases of dynamic variabilities. 

The proposed approach captures interactions among various abstractions that impact algorithm selection (e.g., dataset, prediction type, and outcomes such as performance) as these elements change over time. For example, the data sample size may increase, resulting in the selection of different algorithms. Further, the prediction category may change as additional quantitative data become available, new important features are discovered and can be added to the dataset, or changes in outcomes (e.g., accuracy) may result in changes in the models (e.g., the drift problem \cite{lu2018learning}. In more dynamic settings, according to Hummer et al. \cite{hummer2019modelops}, while in the classical application lifecycle, new builds are triggered by code base changes, in the AI application lifecycle, new builds could be triggered by data or code changes, which may activate a re-training process or require the replacement of the current ML model for a new one.

Research aiming to develop automated pipelines has faced several obstacles to efficiently incorporating ML models in production, especially when it comes to monitoring and adapting to changes that occur over time \cite{symeonidis2022mlops, karmaker2021automl}. 

Indeed, the ML operational environment can be affected by changes over time. In the ML modeling process, for example, data sets may evolve, or the model performance may deteriorate, and these changes may require changing the selected model algorithm. This means that in a changing environment where ML applications operate, constant model retraining may be required to ensure the ML application remains with appropriate performance, quality, and efficiency. This scenario indicates a need for constant monitoring of ML model applications after model deployment, enabling detection of outliers and data drifts, assessing model performance and metrics of incoming data, and supporting adaptation triggered by the monitored changes.

In the ML project lifecycles proposed in the literature, such as the Microsoft ML lifecycle \cite{amershi2019software}, the workflow takes into consideration the continuous monitoring stage after the model deployment stage, enabling to loop back to any of the previous stages in different ways to cope with the needed changes and reconfiguration.  

For example, a machine learning model is based on data, and since data is continuously changing, the model should be retrained to ensure a continuous improvement that will result in an efficient ML application outcome \cite{symeonidis2022mlops, klaise2020monitoring}. 
According to \cite{klaise2020monitoring}, ensuring high-quality ML services requires additional procedures beyond the deployment stage of the ML lifecycle, including areas such as data monitoring and model performance. There are also situations in which concept drifts, such as differences between training and test sets, may affect the predictive performance and result in adaptive changes. 

We also assess other possibilities for the same application based on changes that can occur at design or runtime, showing how these changes could drive adaptations in the ML model selection. 

\begin{figure*}[!ht]
	\centering
    \includegraphics[width=\linewidth]{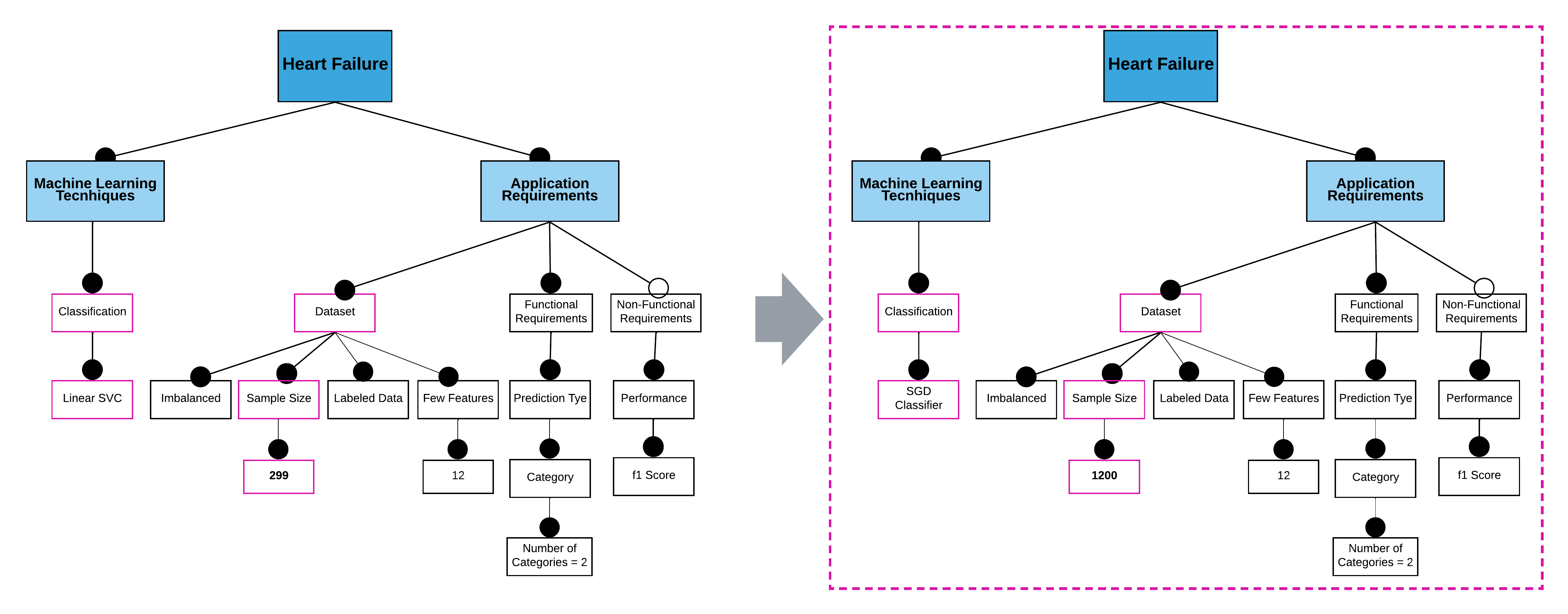}
	\centering
	\caption{Feature model representing an adaptation triggered by changes in the application dataset.}
	\label{fig:Adaptation_SampleSize}
\end{figure*}

Figure \ref{fig:Adaptation_SampleSize} illustrates an adaptation triggered by changes in the application dataset. Assume that over time, the volume of data has considerably increased to more than 100K entries. According to the Scikit-Learn heuristics, for a dataset with more than 100K entries, the SGD classifier is the algorithm recommended. Therefore, a new feature diagram should be instantiated to represent the application after this adaptation. For instance, after selecting a new feature to describe the application dataset, the system replaces the Linear SVC technique with the SGD classifier.

\begin{figure*}[!ht]
	\centering
    \includegraphics[width=\linewidth]{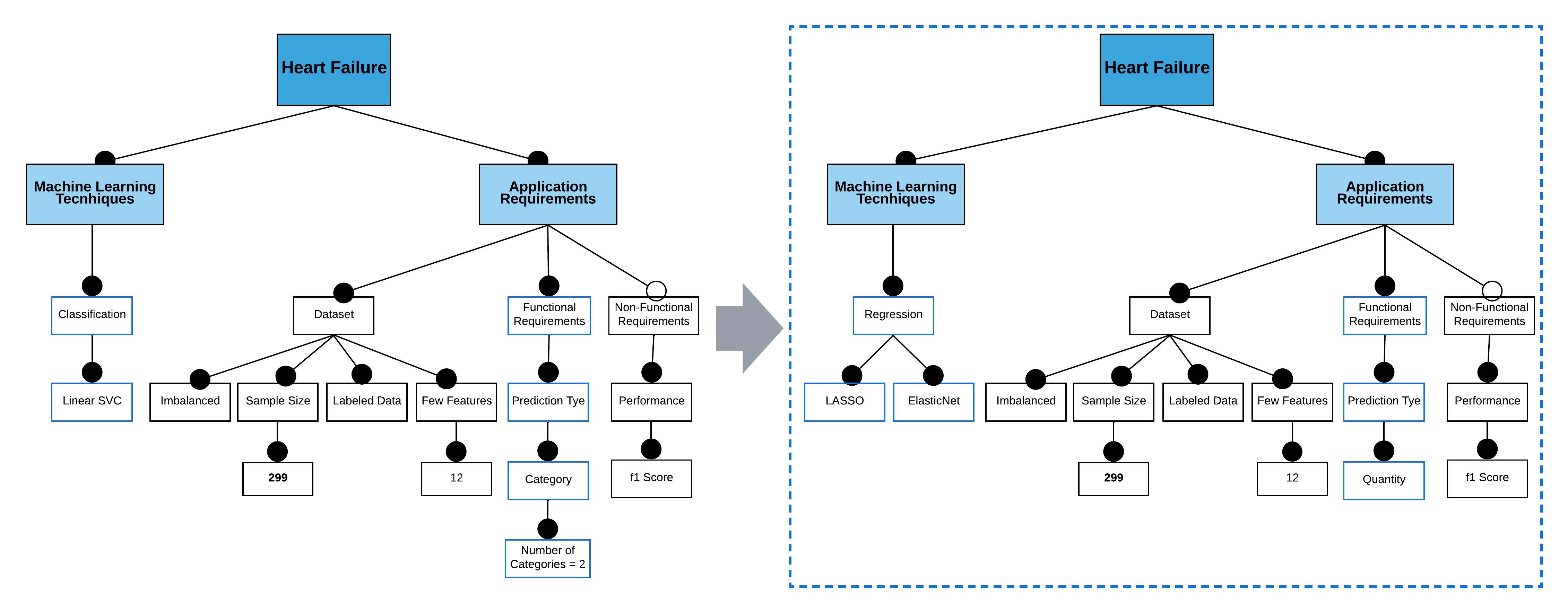}
	\centering
	\caption{Feature model representing an adaptation triggered by changes related to design decisions.}
	\label{fig:Adaptation_death_time}
\end{figure*}

Figure \ref{fig:Adaptation_death_time} illustrates an adaptation that may be triggered by changes related to design decisions, such as functional requirements. Supposing the data volume is the same as the first iteration, but the application goal changed. Instead of predicting whether a patient would die within 130 days, designers decided to predict how many days the patient would die. In this case, the application's main purpose changed from predicting a category to predicting a quantity. When predicting quantity, the Scikit-Learn heuristics recommends regression algorithms instead of classification. In such a case, considering the same dataset configuration, the LASSO or Elasticnet algorithms would be used. This figure illustrates the structural changes made to the feature model to represent the application after this adaptation.

\section{Conclusion and Future Work}
In this section, we present our conclusions and discuss open challenges that can be tackled in future work.

\subsection{Conclusions}
In this paper, we present a variability-aware ML model selection approach. The approach identifies and models factors that capture the variability of ML model selection and their dependencies. It shows how the models can be instantiated to capture a specific model selection heuristics (i.e., Scikit-Learn), and describes an experimental case study based on performance metrics to demonstrate the applicability of the approach. The presented ML model selection phase can benefit both designers and practitioners as it can lead to cost and time savings and make the selection accessible to non-expert users.

The work advances the state of the art in developing methods to support the design and automation of ML model selection in ML application development. In contrast with the current approaches, which are often ad hoc and informal, the proposed approach can be seen as a step towards the provision of a more explicit, systematic, transparent, interpretable and automated basis for model selection. 

\subsection{Future Work}
This work can be extended in several ways, which are described in the following paragraphs. 

\subsubsection{Increasing the model coverage of the ML model selection process}

Heuristics presented by the Scikit-Learn flowchart to select a model does not address many methods that are already supported by the Scikit-Learn library. As future work, we can explore additional documentation associated with this library to identify heuristics that are not still covered by the Scikit-Learn official flowchart.
For example, Least-angle regression (LARS) is a regression algorithm for high-dimensional data that is recommended for datasets for which the number of features is significantly greater than the number of samples \cite{pedregosa2011scikit}. 

We can also include additional heuristics to increase the coverage of ML model selection.
For example, in terms of LASSO, it is stated that it "prefers solutions with fewer non-zero coefficients, effectively reducing the number of features upon which the given solution is dependent." 
In other words, LASSO works better with a small set of features. Therefore, in case there are many data entries for which specific feature values are null or zero, that is, features that are useless, these features can be removed to decrease the number of dataset features.

In some cases, more detailed coverage can be provided. Scikit-Learn supports the use of eight Ensemble Classifiers \cite{scikit-api}, but its heuristics flowchart does not distinguish selection among the Ensemble Classifiers. 

In addition, although we have included software quality attributes represented as 'Working' (e.g.,  'Accurate') in a feature diagram, in future work, constraints can be explored that describe and limit the relation between the quality attributes and the modeling technique feature based on the feature model.

\subsubsection{Exploring other heuristic perspectives}
Other entities are making a growing effort to try to find ways to make model selection easier. For example, Microsoft proposed guidance to select models using specific heuristics \cite{microsoft-heuristic}. In future work, we can explore the proposed Microsoft ML algorithm cheat sheet and integrate its heuristics into our approach.

We can also conduct qualitative research with ML engineers and data scientists to identify the heuristics they actually use while selecting machine learning models. 

\subsubsection{Heuristic-guided parallel search}
 Leaning et al. \cite{leenings2021photonai}  proposed an approach to run several methods in parallel until the best fit is identified. Even when performing a parallel execution, there are several options for creating algorithm sets. Also, depending on the dataset size, performing several processes can be costly performance-wise and time-consuming. This approach could be enhanced by using heuristics to select the algorithms that will compose the search set, limiting the search space and increasing assertiveness. 
 
 This solution of running several in parallel can also help in situations where we do not have heuristics for specific subsets of algorithms. For example, in the case of Scikit-Learn, heuristics were provided to reach a group of algorithms, the Ensemble Classifiers. However, no heuristics were provided to find an algorithm from that group that best fits the problem. 

\subsubsection{Conducting additional case studies} 
Our work can also be extended by selecting additional experiments described in papers that use one or more ML classification modeling techniques and developing additional case studies that show the applicability of the proposed approach. 
  
 \subsubsection{Exploring heuristics that allow us to go beyond the selection of algorithms}
 In this case, it is possible to explore heuristics oriented towards parameter selection and automatic configuration of algorithms. The configuration of algorithms is a topic that is considered in AutoML.  
 
 \subsubsection{Exploring the relation between different quality criteria and ML algorithm selection}
 It is known that the importance of software qualities varies according to the application domain. 

 Future work can investigate additional software qualities or non-functional requirements such as fairness and explore constraints that describe and limit the relation between quality and modeling techniques features based on this feature model.
 
\section*{Acknowledgment}
The Natural Sciences and Engineering Research Council of Canada (NSERC) and the Centre for Community Mapping (COMAP) supported this work.

\bibliographystyle{IEEEtran}
\bibliography{main}

\vspace{12pt}

\end{document}